\RequirePackage{fix-cm}
\documentclass[smallextended]{svjour3}       
\smartqed  
\usepackage{graphicx}
\begin{document}

\title{Determination of the $\mathbf{\Sigma(1385)^{0}/\Lambda(1405)}$ ratio in p+p collisions at 3.5 GeV}

\author{E. Epple and L. Fabbietti for the HADES collaboration} 


\institute{ Laura Fabbietti \at
              Excellence Cluster Universe \\
              Tel.: +49-89-35831 7136\\
              \email{laura.fabbietti@ph.tum.de}           
}

\date{Received: date / Accepted: date}
\maketitle

\begin{centering}
G.~Agakishiev$^{6}$, A.~Balanda$^{3}$,
D.~Belver$^{17}$, A.V.~Belyaev$^{6}$, A.~Blanco$^{2}$, M.~B\"{o}hmer$^{9}$, J.~L.~Boyard$^{15}$, P.~Cabanelas$^{17}$, E.~Castro$^{17}$, J.C.~Chen $^{8}$
% ...

\vspace*{0.3cm}
\normalsize{{(HADES collaboration)\\}}
\noindent \textit{$^{1}$Istituto Nazionale di Fisica Nucleare - Laboratori Nazionali del Sud, 95125~Catania, Italy}\\
\textit{$^{2}$LIP-Laborat\'{o}rio de Instrumenta\c{c}\~{a}o e F\'{\i}sica Experimental de Part\'{\i}culas , 3004-516~Coimbra, Portugal}\\
\textit{$^{3}$Smoluchowski Institute of Physics, Jagiellonian University of Cracow, 30-059~Krak\'{o}w, Poland}\\
\textit{$^{4}$GSI Helmholtzzentrum f\"{u}r Schwerionenforschung GmbH, 64291~Darmstadt, Germany}\\
\textit{$^{5}$Institut f\"{u}r Strahlenphysik, Forschungszentrum Dresden-Rossendorf, 01314~Dresden, Germany}\\
\textit{$^{6}$Joint Institute of Nuclear Research, 141980~Dubna, Russia}\\
\textit{$^{7}$Institut f\"{u}r Kernphysik, Goethe-Universit\"{a}t, 60438 ~Frankfurt, Germany}\\
\textit{$^{8}$Excellence Cluster 'Origin and Structure of the Universe' , 85748~Garching, Germany}\\
\textit{$^{9}$Physik Department E12, Technische Universit\"{a}t M\"{u}nchen, 85748~Garching, Germany}\\
\textit{$^{10}$II.Physikalisches Institut, Justus Liebig Universit\"{a}t Giessen, 35392~Giessen, Germany}\\
\textit{$^{11}$Institute for Nuclear Research, Russian Academy of Science, 117312~Moscow, Russia}\\
\textit{$^{12}$Department of Physics, University of Cyprus, 1678~Nicosia, Cyprus}\\
\textit{$^{13}$Frederick University, 1036 Nicosia, Cyprus}\\
\textit{$^{14}$Institut de Physique Nucl\'{e}aire (UMR 8608), CNRS/IN2P3 - Universit\'{e} Paris Sud, F-91406~Orsay Cedex, France}\\
\textit{$^{16}$Nuclear Physics Institute, Academy of Sciences of Czech Republic, 25068~Rez, Czech Republic}\\
\textit{$^{17}$Departamento de F\'{\i}sica de Part\'{\i}culas, Univ. de Santiago de Compostela, 15706~Santiago de Compostela, Spain}\\

\vspace*{0.1cm}
\noindent\textit{$^{a}$ also at Lawrence Berkeley National Laboratory, ~Berkeley, USA}\\
\textit{$^{b}$ also at ISEC Coimbra, ~Coimbra, Portugal}\\
\textit{$^{c}$ also at ExtreMe Matter Institute EMMI, 64291~Darmstadt, Germany}\\
\textit{$^{d}$ also at Technische Univesit\"{a}t Darmstadt, ~Darmstadt, Germany}\\
\textit{$^{e}$ also at Technische Universit\"{a}t Dresden, 01062~Dresden, Germany}\\
\textit{$^{f}$ also at Dipartimento di Fisica, Universit\`{a} di Milano, 20133~Milano, Italy}\\
\textit{$^{g}$ also at Dipartimento di Fisica Generale and INFN, Universit\`{a} di Torino, 10125 Torino, Italy}

\maketitle
\begin{abstract}
The aim of the present analysis is to determine the relative production cross sections of the $\Lambda$(1405) and $\Sigma(1385)^{0}$ resonances in p+p collisions at E$_{kin}$=3.5 GeV measured with HADES.
Upper and lower limits have been determined for the ratio $\sigma_{(\Sigma(1385)^{0}+p+K^{+})}/\sigma_{(\Lambda(1405)+p+K^{+})}=0.76_{-0.26}^{+0.54}$.
The knowledge of this ratio is an essential input for the analysis of the decay $\Lambda(1405)\rightarrow\Sigma^{\pm}\pi^{\mp}$,
where an unambiguous separation of the $\Lambda$(1405) and $\Sigma(1385)^{0}$ signals is not possible.
\end{abstract}
\end{centering}
\section{Introduction}
\label{intro}
Among the baryon resonances listed in the Particle Data Group tables \cite{pdg} the $\Lambda$(1405) exhibits exceptional characteristics. 
Indeed, effective field theories \cite{LATheory} describe it as 
a superposition of a $\Sigma\pi$ resonance and a $\bar{K}N$ bound state contributing to the resonance 
formation. It is assumed, moreover, that these pole contributions are populated with different strengths depending on the entrance reaction. 
Thus, it is rather important to access the spectral shape of the $\Lambda$(1405) in different entrance channels.\\
The data discussed here have been collected with the \textbf{H}igh \textbf{A}cceptance \textbf{D}i-\textbf{E}lectron \textbf{S}pectrometer (HADES) \cite{Technical} at GSI.
In the measured p+p reaction (E$_{kin}$=3.5 GeV, fixed target) the $\Lambda(1405)$ and the $\Sigma(1385)^{0}$ resonances are produced together with a proton and a K$^{+}$ meson. The two charged decay channels of the $\Lambda(1405)$ resonance, $\Sigma^{\pm}\pi^{\mp}$, have been also analyzed for this data set and the results are reported in \cite{Johannes}. There, a separation of $\Lambda(1405)$ and $\Sigma(1385)^{0}$ signals is not possible. Hence, an external reference is needed to determine their relative yields.
This reference is delivered by the analysis of the $\Lambda(1405)$ decay channel $\Sigma^{0}\pi^{0}$. Indeed, the two resonances
can be partially disentangled in this decay channel, since the decay of the $\Sigma(1385)^{0}$ into $\Sigma^{0}\pi^{0}$ is forbidden, and limits for the ratio $\sigma_{(\Sigma(1385)^{0}+p+K^{+})}/\sigma_{(\Lambda(1405)+p+K^{+})}$ can be estimated.
\section{Analysis Procedure}
\label{sec:1}
The analysis presented here aims to reconstruct the reaction:\\
$p+p\rightarrow p+K^{+}+Y$, where $Y=\Lambda(1405)$ and $Y$=$\Sigma(1385)^{0}$ are of interest. When detecting the proton and the $K^{+}$ meson the spectral shape of the resonances can be reconstructed by the missing mass $MM_{(pK^{+})}$. To further select the data sample and at least partially disentangle these
two resonances, their charged decay products have to be detected as well. The analyzed decays are 
$\Lambda(1405)\rightarrow\Sigma^{0}\pi^{0}\rightarrow(\Lambda\gamma)\pi^{0}$ and $\Sigma(1385)^{0}\rightarrow\Lambda\pi^{0}$. The second step in the analysis consists in selecting a $\Lambda$ hyperon in the final state. The decay $\Lambda\rightarrow p\pi^{-}$   
(BR=63.9 \%) is considered, where the $\Lambda$ candidates have been selected by applying the following cuts:
(1) distance between the proton and pion track (d$_{p-\pi^{-}}$ $<$ 18mm), 
(2) the distance of closest approach of the $\Lambda$ to the primary vertex (DCA$_{\Lambda}$ $<$ 23mm), 
(3) the constraint (DCA$_{p}$ $<$ DCA$_{\pi^{-}}$) and (4) a cut on the $p \pi^{-}$ invariant mass ($1106\, <\, M_{(p,\pi^{-})}\,<\, 1122 \,MeV/c^{2}$). 
The obtained missing mass spectrum of a proton and a $K^{+}$ meson shown in Fig.~1 consists of two different sets of data. 
The left panel of Fig. ~1 shows the 'HADES data set', where all charged particles were detected within HADES \cite{Technical}.  The right panel of Fig. ~1 shows the 'WALL data set', where the proton from the $\Lambda$ decay is detected in the Forward Wall\footnote[1]{An external hodoscope placed $7\,\mathrm{m}$ downstream of the target with a time resolution of $500-700\, \mathrm{ns}$.} and the other three particles in HADES.
These two data sets can be analyzed independently and provide complementary information as they cover different regions of the phase 
space (a resolution of $\sigma=\,4\,\mathrm{MeV/c}^{2}$ and $2\, \mathrm{MeV/c}^{2}$ for the reconstructed $\Lambda$ mass is obtained for the 'WALL' and 'HADES data set', respectively  \cite{EppleMeson}). 
Both data sets show a similar resolution in the missing mass $MM_{(p,K^{+})}$ ($ \sigma=\,24\, (18)\, \mathrm{MeV/c}^{2}$ and $\sigma=\,21\, (15)\, \mathrm{MeV/c}^{2}$ at the $\Lambda$ and $\Sigma^{0}$ pole positions, respectively).
%
%
\begin{figure}
\begin{flushleft}
\begin{minipage}[h]{0.45\textwidth}
  \includegraphics[width=1\textwidth]{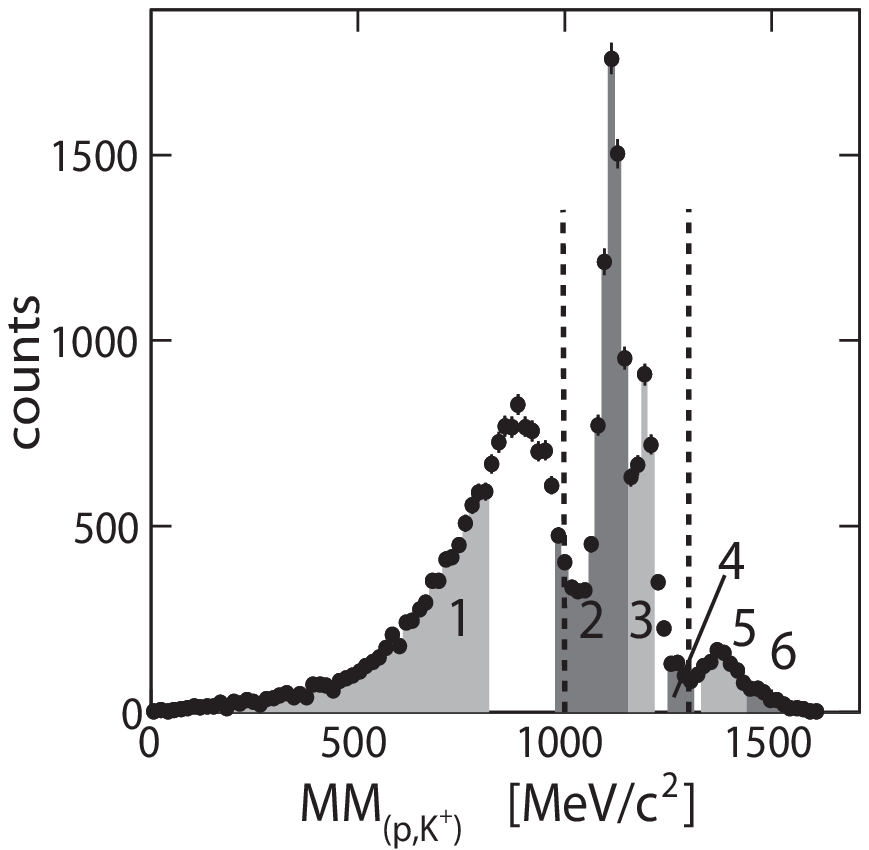}
\end{minipage}
\hspace{0.5cm}
\begin{minipage}[h]{0.45\textwidth}
  \includegraphics[width=1\textwidth]{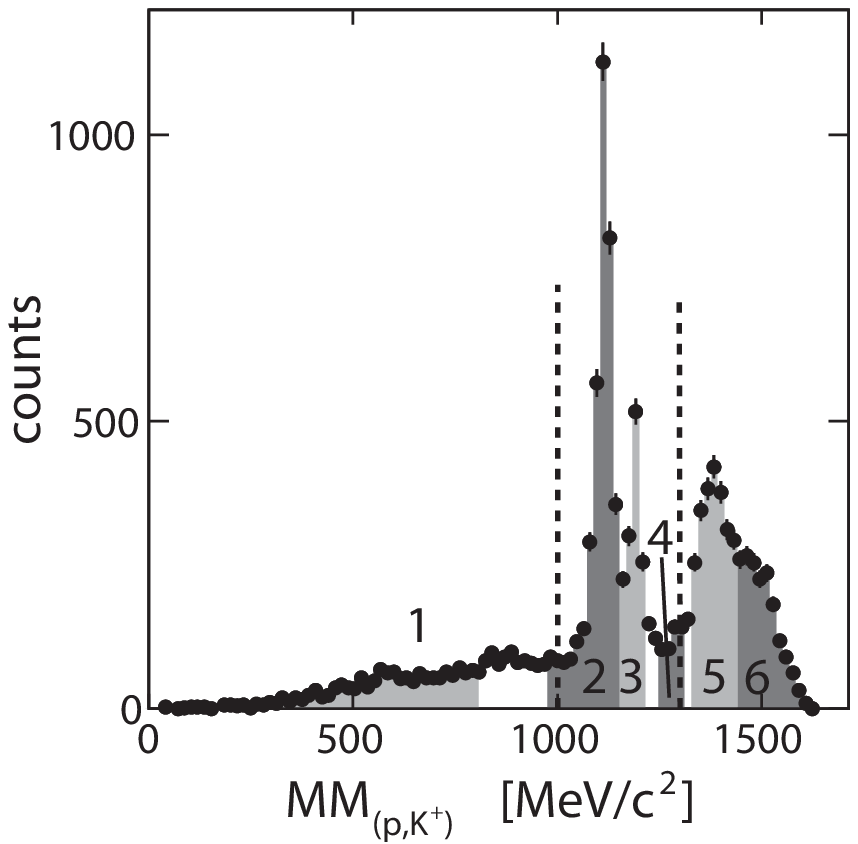}
\end{minipage}
\caption{The missing mass $MM_{(p,K^{+})}$ for the 'HADES data set' (left) and the 'WALL data set' (right). The numbered areas show the six bins in the missing mass $MM_{(p,K^{+},p,\pi^{-})}$ distribution
that have been simultaneously fitted. The black dashed lines separates the three intervals in which the validation of the misidentification background has been cross-checked independently.}
\end{flushleft}
\label{fig1}
\end{figure}
The possible reactions with a final state p+K$^{+}$+X contributing to the missing mass $MM_{(p,K^{+})}$ distribution are listed in table ~\ref{tab:1}. Besides several reactions involving strangeness production, the total measured yield also contains some misidentification background by $p$ and $\pi^{+}$ identified as $K^{+}$ (cf. \cite{SigmaResonance}, section IV). 
%
%
It is however possible to model this background precisely by a side-band analysis of the reconstructed K$^{+}$ mass spectra. 
An exhaustive description of this data driven side-band method is available in \cite{SigmaResonance}. 
To determine the exact yield of the misidentification background in the analyzed spectra the side-band data sample has to be 
compared to the measured sample. In this analysis we choose to compare the spectra by means of the $p-\pi^{-}$ invariant mass, as illustrated in Fig. 1 of \cite{EppleMeson}, where the $p-\pi^{-}$ invariant mass is displayed under the condition $MM_{(p,K^{+})}>$ 0 MeV/c$^{2}$. 
The side-band data sample was additionally compared to two further $p-\pi^{-}$ invariant mass spectra, applying the following conditions:  $MM_{(p,K^{+})}>$ 1000 MeV/c$^{2}$ and $MM_{(p,K^{+})}>$ 1300 MeV/c$^{2}$. These three intervals are defined between the vertical dashed lines, shown in Fig. ~1.
In this way, the yield of the side-band background could be cross-checked in one observable, independently for different data sub-samples selected by means of $MM_{(pK^{+})}$.
\begin{figure}\label{Fig2}
\includegraphics[width=1\textwidth]{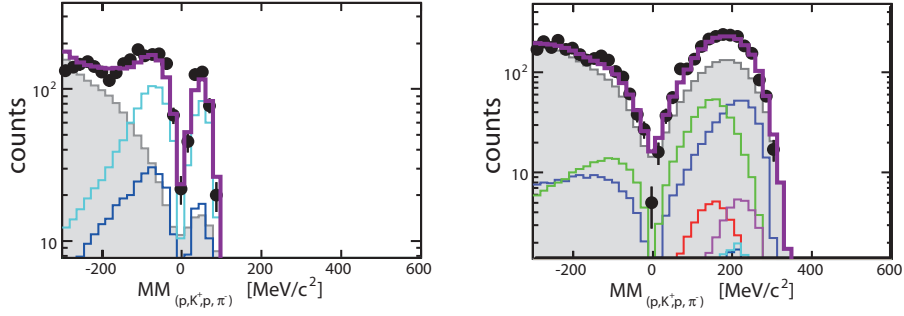}
 \caption{An example of a fit result in two of the 14 fitted bins of the missing mass $MM_{(p,\pi^{-},p,K^{+})}$ for the 'WALL data set'. On the left the third interval displayed in Fig. 1 and on the right the fifth interval is displayed. The experimental points are shown in black, the sum of all fitted contributions is given in purple. The gray area displays the misidentification background. The colored curves correspond to the different simulated channels listed in table \ref{tab:1}.}
\end{figure} 
The precision of the side-band method is limited by the fact that the $p$- and $\pi^{+}$-misidentification samples are obtained by 
a side-band selection on the reconstructed kaon mass in two intervals rather distant from the nominal $K^{+}$ mass. Even by doing so, some real kaons might be in the two background samples.
This contribution is responsible for the fact that the side-band background underneath the $M_{(p,\pi^{-})}$ signal is not flat but shows a small $\Lambda(1116)$ signal \cite{EppleMeson}. The strength of physical signal in the side-band statistic can be evaluated by filtering simulations 
through the side-band analysis. In this way, one can determine the fraction of signal in the side-band that fulfills all selection criteria. 
In a later step of the analysis this simulated signal is subtracted from the misidentification background.
%
\begin{table}
\caption{Shown is a typical outcome of the fit. The relative scaling factors of the simulated channels p+p$\rightarrow$X  have been normalized to the $\Lambda K^+p$ yield. 
}
\label{tab:1}       
\begin{tabular}{llll}
\hline\noalign{\smallskip}
Number & Color-code & Final state X & Relative scaling factors \\
\noalign{\smallskip}\hline\noalign{\smallskip}
1) & violet     & $\Lambda(1405)K^+p$     & 0.266 \\
2) & light green& $\Sigma(1385)^{0}K^{+}p$& 0.14  \\
3) & dark green & $\Lambda(1520)K^+p$     & 0.257 \\
4) & dark blue  & $\Lambda K^+p$          & 1\\
5) & red        & $\Lambda K^+p \pi^0$    & 0.031 \\ 
6) & cyan       & $\Sigma^0K^+p$          & 0.348 \\
7) & pink       & $\Sigma^0K^+p\pi^0$     & 0.023 \\
8) & beige      & $\Sigma^+K^+p\pi^-$     & 0.008 \\ 
\noalign{\smallskip}\hline
\end{tabular}
\end{table} 
%
The reactions given in table \ref{tab:1} were each simulated with the same statistics of events. The final states $\Lambda$/$\Sigma^{0} K^+p\pi^-\pi^+$ and $\Sigma^+K^+p\pi^-\pi^0$ have not been included since studies based on simulations showed that their contribution to the measured yield is not significant.
Each channel has been simulated to determine the corresponding shape in the $MM_{(pK^{+})}$ distribution. 
Their relative yield has been extracted by fitting the scaled sum of all channels to the experimental distribution.
Six intervals have been defined in the missing mass $MM_{(p,K^{+})}$ for the two data samples separately, as shown in Fig.~1
by the gray areas. For each of the 12 intervals an experimental distribution in $MM_{(p,\pi^{-},p,K^{+})}$ is obtained and later 
fitted with a function composed of the sum of all the simulated contributions listed in table\,\ref{tab:1} each multiplied by a fit parameter and the misidentification background, of which the yield is quantitatively determined by the side-band method.
%
%
%
During the fitting procedure the yield of $\Lambda(1520)$ has been by trend underestimated. Therefore the simulations could not describe the $MM_{(p,K^{+})}$ spectrum in the mass region around 1520 MeV/c$^{2}$. 
An additional distribution corresponding to the interval of 1450 $< MM_{(p,K^{+})}<$ 1620 MeV/c$^{2}$ for both data sets has been included in the fit to solve this problem.
In total 14 experimental distributions were fitted simultaneously.\\
In Fig. 2 the experimental distributions of two out of the 14 bins are shown together with a result of the fit.
%
The following constraints have been applied during the fit :
$\sigma_{(\Lambda K^+p \pi^0)}$ and $\sigma_{(\Sigma^0K^+p\pi^0)}$ $<$ $\sigma_{(\Lambda K^+p)}$. A further constraint has been obtained from the analysis described in \cite{Johannes}. 
Here, the contribution from the final state $\Sigma^+K^+p\pi^-$ appears as a rather flat distribution under  the $\Lambda(1405)$ resonance signal in the $MM_{(pK^+)}$ spectrum. 
The ratio of the resonance signal to the flat contribution has been determined. It results in the constraint $\sigma_{(\Sigma^+K^+p\pi^-)}$ $<$ 0.3 $\cdot$ ($\sigma_{(\Lambda(1405)K^+p)}$+$\sigma_{(\Sigma(1385)^{0}K^{+}p)}$).
In order to describe both data sets simultaneously with the same simulations it has been necessary to include a non-isotropic angular distribution for the simulation of the channel $\Lambda K^+p$.
The reason for this is that the $\Lambda$(1116) is produced with an angular anisotropy in p+p collisions, as discussed in
\cite{Wissermann}. The simulations were filtered with the $\Lambda$ angular distribution in the p-p 
CM system extracted at a beam kinetic energy of 2.4 GeV \cite{Wissermann}. The experimental data could have been described much 
better with this modified simulation.
%
%
\begin{figure}\label{Fig3}
\begin{flushleft}
\begin{minipage}[h]{0.45\textwidth}
  \includegraphics[width=1.03\textwidth]{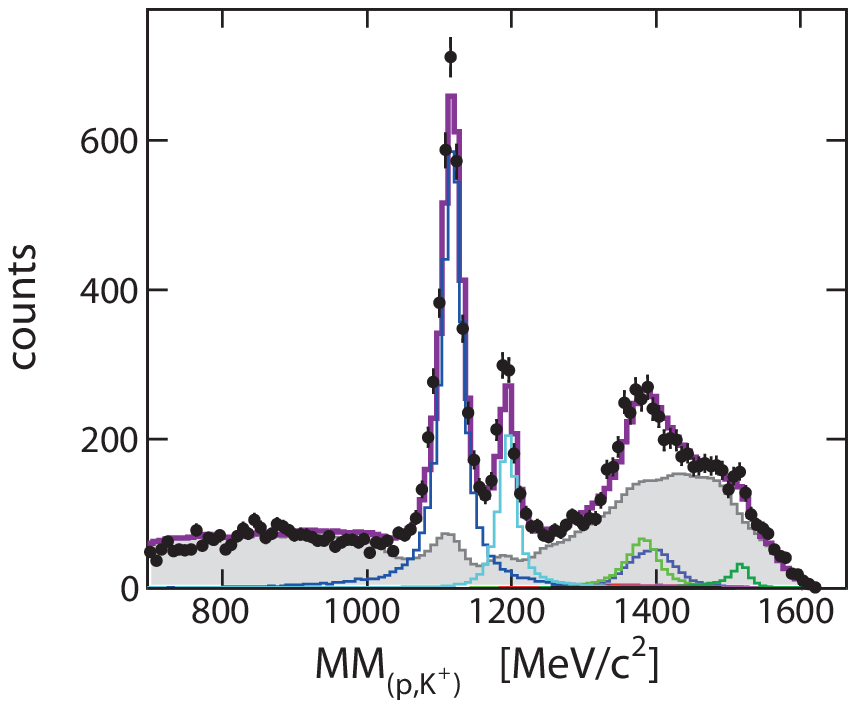}
\end{minipage}
\hspace{0.3cm}
\begin{minipage}[h]{0.45\textwidth}
  \includegraphics[width=1.1\textwidth]{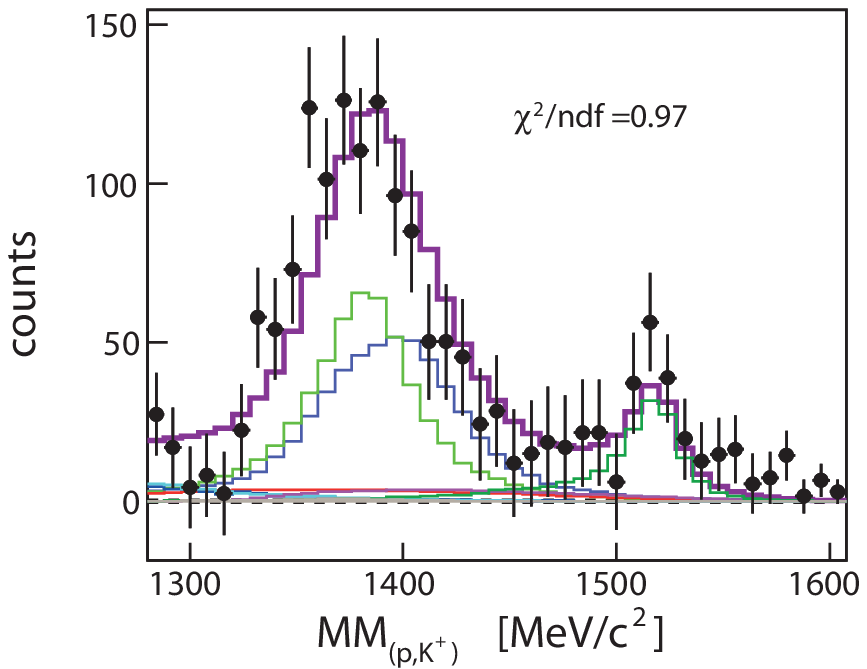}
\end{minipage}
\caption{Fitted distributions shown together with the experimental distribution of $MM_{(p,K^{+})}$ (left) and for the resonance mass area (right), after subtracting the misidentification background. The experimental points are black, the sum of all fitted contributions is shown in purple. The gray area is the misidentification background. The colored curves correspond to the different simulated channels listed in table \ref{tab:1}.}
\end{flushleft}
\end{figure}
Recent data from the same experiment \cite{SigmaResonance} have shown, that also the $\Sigma(1385)^{+}$ resonance is produced 
with an angular anisotropy. Thus in a second step the simulations for the final state $\Sigma(1385)^{0}K^{+}p$ were weighted with 
the angular distribution of the $\Sigma(1385)^{+}$ in the pp CM frame reported in \cite{SigmaResonance} to study the impact on 
the fit parameters. A final result is shown in Fig.~3, where the $MM_{(p,K^{+})}$ experimental distribution is shown together 
with the contributions of the misidentification background and the simulated channels for the whole range (left panel) and for 
the range, where the $\Lambda(1405)$ appears after subtraction of the misidentification background (right panel).
The obtained $\chi^{2}$ per degree of freedom of the simultaneous fit to the HADES and WALL data sets amounts to 2.1. This 
corresponds to the scaling factors for the different channels listed in table\,\ref{tab:1}. 
Systematic studies were carried out by varying the track cuts and the invariant mass interval for the $\Lambda(1116)$ 
(1), (2) and (4) by $\pm$10 and $\pm$ 20 \% and varying the yield of the misidentification background by 4 to 6 \%. Furthermore, the stability of the fit has been tested by assuming either a flat angular distribution for the $\Sigma(1385)^{0}$ production or
the anisotropic one extracted from the $\Sigma(1385)^{+}$ production.
Within these variations we have obtained a distribution of the $\sigma_{\Sigma(1385)^{0}K^{+}p}$/$\sigma_{\Lambda(1405)K^+p}$ ratio of $0.76_{-0.26}^{+0.54}$ (syst.), where the systematic error is the $\sigma$ value of the distribution obtained by the cut variations. 
These values can be compared with the result obtained by the ANKE collaboration at E$_{kin}$=2.8 GeV \cite{ANKE}, where $0.89\pm0.46$ (stat.)$\pm0.5$ (syst.) has been determined as ratio.
The extracted value for the ratio of $\Sigma(1385)^{0}$/$\Lambda(1405)$ = 1.1 has been employed in the analysis described in \cite{Johannes}.
\begin{acknowledgements}
The author gratefully acknowledges support from the TUM Graduate School.
The following funding are acknowledged:
LIP Coimbra, Coimbra (Portugal): PTDC/FIS/113339/2009,
SIP JUC Cracow, Cracow (Poland): NN202286038, NN202198639, 
HZ Dresden-Rossendorf, Dresden
(Germany): BMBF 06DR9059D, TU Muenchen,
Garching (Germany) MLL Muenchen DFG EClust:
153 VH-NG-330, BMBF 06MT9156 TP5 TP6, GSI
TMKrue 1012, GSI TMFABI 1012, NPI AS CR,
Rez (Czech Republic): MSMT LC07050, GAASCR
IAA100480803, USC - S. de Compostela, Santiago de
Compostela (Spain): CPAN:CSD2007-00042, Goethe
Univ. Frankfurt (Germany): HA216/EMMI, HIC
for FAIR (LOEWE), BMBF06FY9100I, GSI F\&E01,
CNRS/IN2P3 (France).
\end{acknowledgements}
%
%

\end{document}